\documentclass[10pt]{article}
\usepackage[utf8]{inputenc}
\usepackage[T1]{fontenc}

\usepackage{graphicx}

\usepackage{lipsum}
\usepackage{xcolor}

\usepackage{authblk}
\usepackage[margin=2cm]{geometry}
\usepackage{mathtools, cuted}

\usepackage{bm}
\usepackage{amsfonts}
\usepackage{amsmath}
\usepackage{amssymb}
\usepackage{mathrsfs} 
\usepackage{siunitx}

\usepackage{soul}
\usepackage[normalem]{ulem}

\title{
Reducing Foam Friction with Self Slippery Liquid-Infused Porous Surfaces}
\author[1]{Alexis Commereuc}
\author[1]{Emmanuelle Rio}
\author[1,2]{François Boulogne}
\affil[1]{\small Laboratoire de Physique des Solides, UMR 8502, CNRS, Université Paris-Saclay, 91405 Orsay, France.}
\affil[2]{\small francois.boulogne@cnrs.fr}

\date{\today}
\begin{document}

        \maketitle
        \begin{abstract}
 Acquiring a comprehensive understanding of the interplay between foam friction and surface roughness is essential for achieving precise control over their flow dynamics.
In particular, a major challenge is to reduce friction, which can be achieved with rough surfaces in the situation where a liquid infuses the asperities. In this study, we propose to explore self-infused surfaces.
We first present simple observations to demonstrate the effectiveness of our surface design by recording the motion of a foam puddle on a smooth surface and a self-SLIPS.
To quantify friction reduction, we conduct stress measurements on surfaces moved at a constant velocity.
Finally, we interpret the variation of the friction force with the velocity by a model considering an effective slip length of the surface.
 This research paves the way for a novel approach to mitigate dissipation in liquid foam flows, holding significant implications for reducing energy consumption in conveying foams for industrial processes and various end-use applications.
        \end{abstract}

%
%

\section{Introduction}

Whether it is for simple fluids like water or complex fluids, boundary conditions play a crucial role in understanding and controlling their flow.
In complex fluids, wall slippage is often observed such as for polymer melts, polymer solutions, emulsions, suspensions \cite{Barnes1995}, and also liquid foams \cite{Kraynik1988,Saugey2006,Marze2008}.
Surface texturing has been successfully employed as a means to increase the friction at solid walls for various complex fluids such as emulsions \cite{Nickerson2005} and foams \cite{Khan1988}.
Indeed, to perform rheological measurements of the bulk fluid properties, it is crucial to prevent wall slippage and ensure to apply shear to the bulk of the tested material.
In such case, asperities offer anchoring points for instance for the soap films of liquid foams  \cite{Khan1988,Marchand2020}.
Nevertheless, surface texturing has also the potential to increase slip at solid walls.
An illustration of this phenomenon can be found in nature, such as with the surface of pitcher plants \cite{Wong2011}.
 By introducing a non-miscible fluid in surface textures, surfaces with exceptional slippery properties, known as SLIPS (Slippery Liquid-Infused Porous Surfaces), can be obtained \cite{Lafuma2011, Wong2011, Hardt2022}.
This new class of materials has opened up various applications where such exceptional slippery properties are desirable, such as in anti-icing \cite{Heydarian2021}, anti-biofouling \cite{Epstein2012}, and self-cleaning surfaces \cite{Wong2011,Pakzad2022}.

However, for the specific case of liquid foams, these textured surfaces infused with oil can be problematic.
In fact, a foam placed in contact with an oil layer can carry the oil in the liquid channels \cite{Mensire2017}, ruining the infused surface.
In addition, SLIPS have been recently demonstrated as effective anti-foaming agents \cite{Wong2021}.
Although these findings can be useful in processes where foam is undesired, they also imply that oil-based surfaces cannot be used to reduce foam friction.

Our objective is to produce a surface reducing the friction with respect to a smooth surface and compatible with aqueous foams.
The design of the proposed surface relies on their roughness and their hydrophilicity such that they are infused by a small amount of liquid extracted from the liquid foam itself, which makes these surfaces compatible with the foam.
Hence, these self slippery liquid-infused surfaces or self-SLIPS are expected to enable friction reduction without the need for additional phase.
We start by showing that such surfaces are effectively reducing the foam friction with simple observations on inclined surfaces.
Then, we perform measurements with controlled foam properties and at a constant surface velocity to record the friction stress and quantify the stress reduction  on self-SLIPS.

\begin{figure*}[tbhp]
\centering
\includegraphics[width=.85\linewidth]{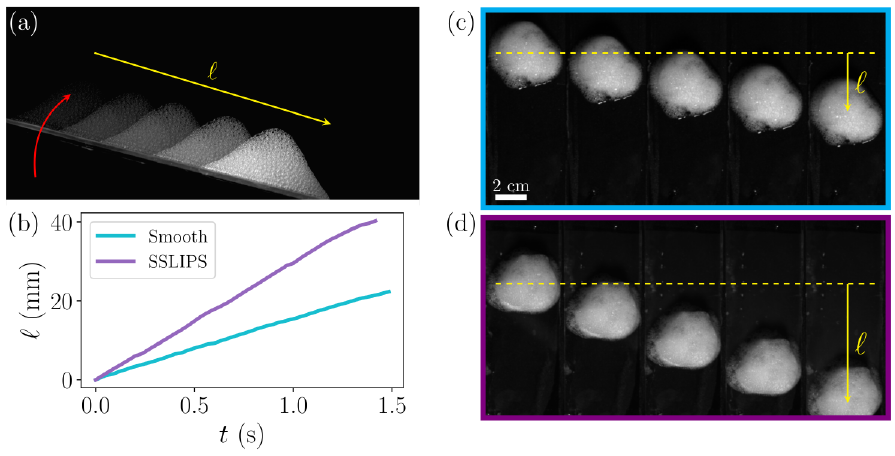}
\caption{(a) Image illustrating the experiment from a side view.
A small amount of foam is put on top of horizontal smooth and textured surfaces.
Surfaces are lifted up to 68$^\circ$ from horizontal to create a gravity flow.
(b) Evolution of the position of the foam along the surfaces with respect to time. In blue for the smooth surface and in purple for the textured one. The velocity is higher in the second case, illustrating the SSLIPS effect.
(c-d) Series of photographs illustrating the foam flowing on surfaces. The colors match the legend of the plot. The first one correspond to $t=0$~s, and the yellow dashed line represent the origin for the $\ell$ distance travelled by the foam.
The time step between consecutive images is $0.5$~s.
The movie is provided in SI.
}
\label{fig:simple_exp}
\end{figure*}

%
%

\section{Foam slippage on textured surfaces}

\subsection{Preparation of textured surfaces}

Textured surfaces are produced in two steps.
First, a counter mold  is made by spin coating SU8 resist  (2010, 2025, 2100 Kayaku AdvancedMaterial) on a silicon wafer (Eloïse 6") to produce a uniform thickness $b$.
The coating is pre-baked at 90°C during two minutes and then, exposed to UV light with a mask corresponding to a square lattice of square holes by using optical lithography (MicroWriter ML 3s, Durham Magneto Optics Ltd). The uncured resist is removed with MicroChem’s SU8 developer to get the final counter mold.
The final dimensions of the patterns are 12~cm by 5.5~cm.
Second, we prepare polydimethylsiloxane elastomer (PDMS, Silgard 184, Dow Corning) with a 10:1 ratio between the monomer and the cross-linker.
The mixture is degassed, poured on the counter mold, and cured.
The PDMS layer is peeled off the surface and chemically bonded on a microscope glass slide ($12\times5$~cm$^2$, Ted Pella) by a plasma treatment.
The excess of PDMS is trimmed at the edges of the glass sheet with a razor blade.
The characteristics of the pattern are checked with an optical profiler (Ametek Taylor Hobson CCI HD).


The geometry of our textured surfaces consists in pillars of height $b$ organized on a square lattice.
Pillars have a square cross-section of width $a$ and the edge-to-edge distance between pillars is also $a$.
A first set of surfaces, called isometric, are made with $a=b=60$~$\mu$m and $a=b=100$~$\mu$m, and a second set has a constant lattice parameters $a=60$~$\mu$m and the height $b$ varies between 10 and 80~$\mu$m.

Prior to each experiment, the surface is activated with a plasma generator (Kreisler) and prewetted with the foaming solution to prevent film bursting when the substrate is inserted in the foam.
Then, the probing surface is attached to a force sensor to perform stress measurements.

\subsection{Observations}

Recently, we have shown that the textured surfaces have two different wetting states when they are placed in contact with a liquid foam \cite{Commereuc2024}.
The two relevant length scales are $a$, the size of the asperities and $r_{\rm pb}$, the radius of curvature of the Plateau borders.
When the Laplace pressure of the foam is higher than the capillary pressure associated with the size $a$ of the asperities, $a/r_{\rm pb} > 1$, we found that the soap films in contact with the surface are sitting between the asperities, leading to zigzag bubble contacts.
In contrast, for the opposite situation $a/r_{\rm pb}< 1$, the bubble footprint on the surface is identical to a smooth surface.
More specifically, the liquid infusing the patterns comes directly from the foam, which is characteristic of what we call a self-SLIPS behavior.
This prevents the foam destruction by an \textit{ad hoc} liquid added to reduce artificially the friction.
This infused liquid must reduce the foam friction, as it is observed for droplets and liquids on SLIPS \cite{Lafuma2011, Wong2011, Hardt2022}.

To verify this assumption, we perform a simple experiment where two foam puddles of identical composition and volumes are placed respectively on a smooth glass surface and a textured surface both pre-soaked with the soap solution.
Surfaces are tilted simultaneously and we quantify the sliding motion  by reporting the center of mass of the foam puddle from the image sequence illustrated in  Fig.~\ref{fig:simple_exp}a.
Our observations indicate that the foam puddle slides down the incline faster for the textured surface (Fig.~\ref{fig:simple_exp}b-d).


To perform more controlled experiments, especially regarding the foam liquid fraction  $\varphi_{\ell}$, we propose a different experimental setup where the surface is inserted in a foam column at equilibrium with its soap solution to prevent liquid drainage in the foam.

%
%


The principle of the experiment is to insert a  surface
of controlled roughness in a dry and monodisperse liquid foam produced in a reservoir \cite{Marchand2020}.
The surface is moved horizontally with a motorized translation stage at a constant velocity $V\in[2,30]$~mm/s  through a dedicated opening on the side of the reservoir.

\subsection{Foam generation}

We prepare a soap solution by diluting a commercial surfactant (Fairy with a concentration in surfactant: 5–15 \%) at a concentration of 10 wt.\% in pure water.
The liquid-vapor surface tension is $\gamma = 24.5 \pm 0.1$ mN/m and the viscosity is $\mu_{\ell} = 1.0 \pm 0.2 $ mPa$\cdot$s at 20~$^\circ$C.

The solution is poured in a transparent container and the liquid height is adjusted to get a controlled distance $z_0$ between the liquid interface and the probe substrate.
Eight identical blunt 32G needles pointing upwards are placed at the bottom of the container.
These needles are connected to a pressure generator (OF1, Elveflow, France) to inject air and generate a monodisperse foam by bubbling air in the foaming solution \cite{Drenckhan2015}.
In our experiments, the bubble radius $R$ is $660 \pm 40$ $\mathrm{\mu m}$.

Once the generated foam reaches the top of the container, the air flow is stopped and a glass lid is placed on the top to prevent evaporation.
A vertical gradient of liquid fraction is established in the foam from the balance between the capillary suction in the liquid films and the gravity drainage.
The liquid fraction profile $\varphi_\ell(z)$ is
\begin{equation}\label{eq:phi_ell}
    \varphi_\ell(z) = \hat\varphi \left( \frac{z}{\ell_c} + \left(\frac{\varphi_c}{\hat\varphi}\right)^{-1/2} \right)^{-2},
\end{equation}
where $z$ is the vertical position from the liquid-foam interface, $\varphi_c = 0.26$ and  $\hat\varphi = \ell_c^2 / R^2\delta^2$ with a geometric constant $\delta = 1.73$ \cite{Cantat2013b,Boulogne2023}.
The liquid fraction is computed from $\varphi_\ell(z_0)$  and varies  between $0.30$ and $2.15 $ $\pm \ 0.02 \ \%$.

Therefore, the monodisperse foam has two parameters: the bubble radius $R$ and the liquid fraction $\varphi_\ell$.
The Plateau borders are characterized by their radius of curvature $r_{\rm pb} = R (\varphi_\ell/0.33)^{1/2}$  {\cite{Cantat2013b}}.

\subsection{Force sensor and stress computation}

The force sensor is made of a capacitive sensor (Fogale Nanotech CS200), which measures the displacement of two parallel beryllium copper (Weber Métaux) blades as described in \cite{Restagno2001}.
The displacement is related to the force by a calibration that consists in using controlled weights in a pan connected to the force sensor by the mean of a nylon fiber and a frictionless pulley.

At a constant velocity, we measure a linear increase of the force with the distance of penetration in the foam for all the tested surfaces and experimental conditions.
For a smooth surface of width $w$, the stress is $\tau_{\rm s} =\frac{1}{2w} \frac{{\rm d} F}{{\rm d} p}$, where $p$ is the penetration distance and $F$ the force.
For the textured surfaces, which are rough on the top side and smooth on the bottom side, the stress exerted on the rough side is $\tau = \frac{1}{w} \frac{{\rm d} F}{{\rm d} p} - \tau_{\rm s}$.

\section{Friction on smooth surfaces}


\begin{figure}[tbhp]
\centering
\includegraphics[width=.6\linewidth]{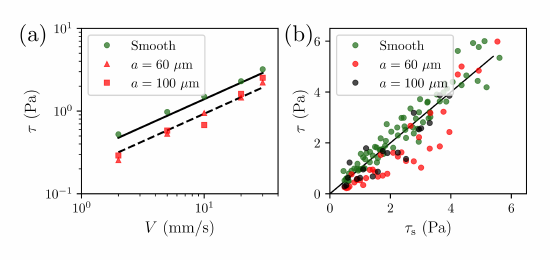}
\caption{
(a) Plot of the measured stress $\tau$ versus the velocity $V$ for $r_{\rm pb} = 92$~$\mu$m.
Greens points are for the smooth surface and red points for the textured surfaces ($a = b$).
The solid black line shows equation \ref{eq:tau_smooth}, which is in good agreement with the data obtained on a smooth surface.
The dashed line represents a power law of 2/3 with respect to the velocity.
(b) Plot of the stress $\tau$ versus the prediction for the smooth surface $\tau_{\rm s}$.
Both the velocity and the liquid fraction effects are well predicted for the smooth case.
\textit{A contrario}, for rough surfaces in red ($a=b=60~\mu$m) and in black ($a=b=100~\mu$m), most of the data are below the prediction.
}
\label{fig:tau_vs_slip}
\end{figure}

In a first set of experiments, we measured the stress exerted by the foam on surfaces with cubic pillars, for one given Plateau border radius of curvature $r_{\rm pb}=92$~$\mu$m  and for different surface velocities $V$.
Fig.~\ref{fig:tau_vs_slip}a shows the results for both a smooth and an isometric textured surface.
On the smooth surface, in green, the stress $\tau$ scales as $V^{2/3}$.
From previous experimental and theoretical studies \cite{Bretherton1961,Hirasaki1985,Cantat2013a,Marchand2020}, foam friction on smooth surfaces is well-described by
\begin{equation}\label{eq:tau_smooth}
    \tau_{\rm s} = k \frac{\gamma}{R} \varphi_\ell^{n} {\rm Ca}^{2/3},
\end{equation}
where $\rm Ca = \eta V/\gamma$ is the capillary number, $ k$ and $n$ are fitting parameters \cite{Raufaste2009,Emile2013}.
From our measurements, we obtain $k=1.94$ and $n = -0.32$, revealing an influence of the liquid fraction as observed previously \cite{Raufaste2009,Emile2013,Marchand2020}.
This model is represented by a black solid line in  Fig.~\ref{fig:tau_vs_slip}a and describes well our data.

In experiments performed in the same conditions with two isometric textured surfaces ($a=b$), the stress is systematically lower than for the smooth surface as presented in Fig.~\ref{fig:tau_vs_slip}a.
This effect can be generalized by systematically varying  the foam liquid fraction, the surface velocity and the surface roughness size for $a/r_{\rm pb} < 1$.
To present the entire data set, in a single plot, we represent in Fig \ref{fig:tau_vs_slip}b the measured stress as a function of the predicted stress for a smooth surface given by Eq. \ref{eq:tau_smooth}.
All the data from the smooth surface are well described by the prediction.
In contrast, data from textured surfaces are below the black line, except at the highest stress values, which correspond to the driest foam near the transition $ a = r_{\rm pb}$.
This emphasizes the existence of a systematic reduction of friction of foams on liquid-infused surfaces.

%
%

\section{Effect of the pillar aspect ratio}

\subsection{Observations}

In a second set of experiments, we maintain constant $a/r_{\rm pb} = 0.65$ and we vary both the height of the pillars $b$ and the surface velocity $V$.
In Fig. \ref{fig:tau_vs_Ca}a, for each pillar height, we schematize a Plateau border, with the soap solution in blue, in contact with the asperities where the proportion is respected.
This shows that the different typical length scales $a$, $b$, $r_{\rm pb}$, have the same order of magnitude.
We thus expect a transition depending on the size $b$.

For each set of parameters, we first measure the stress on a smooth surface.
To quantify the friction reduction, we calculate the ratio $\tau/\tau_{\rm s}$ as a function of $\rm Ca$ in Fig \ref{fig:tau_vs_Ca}b.
The straight black line represents the ratio equal to unity, meaning that there is no drag reduction.
Because no effect of the capillary number is noticeable, the average value of the ratio is presented in dashed line.
Except for the pillar height $b = 10$ $\mu$m, the mean ratio is always below unity.
In Fig. \ref{fig:tau_vs_Ca}c, we report all the average values versus $b$.
A reduction up to 30 $\%$ of the friction is measured for every height $b$ above 10~$\mu$m.

\subsection{Interpretation}

\begin{figure*}[t]
\centering
\includegraphics[width=1\linewidth]{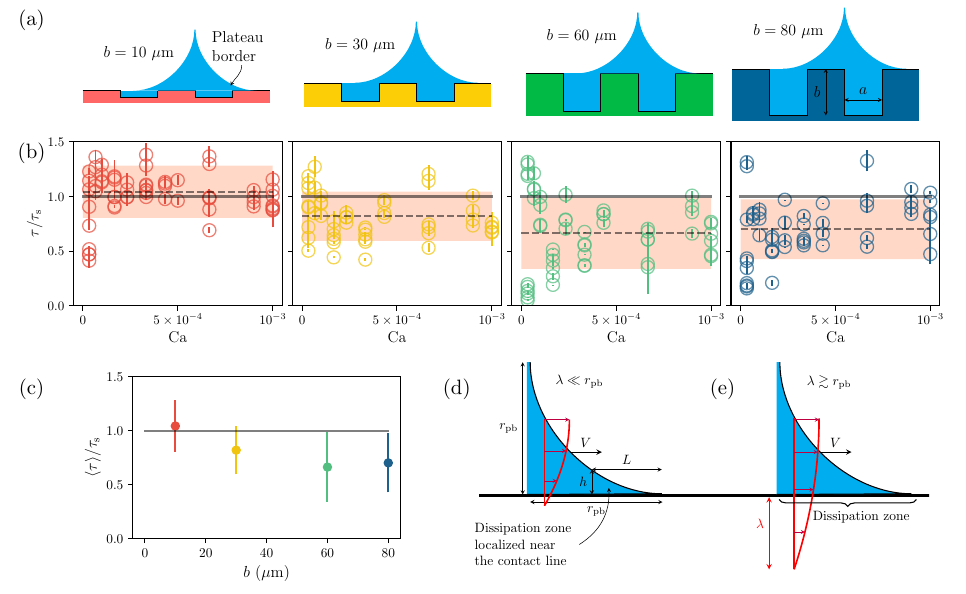}
\caption{(a) Sketch of the textured surfaces for different depth $b$ with a Plateau border at scale for $a/r_{\rm pb} = 0.3$.
(b) Corresponding measured stress ratio $\tau/\tau_{\rm s}$ as a function of the capillary number Ca.
Continuous black lines indicates unity and the dashed lines are the mean values.
The salmon-colored band represents the standard deviation.
(c) Dimensionless stress $\langle \tau \rangle /\tau_{\rm s} $ as a function of the pillars height $b$.
Each point corresponds to the mean value indicated by the dashed line in (b), and the error bars come from the propagation of the standard deviation of both $\tau_{\rm s}$ and $\tau$ calculating their ratio.
(d)  Sketch of a Plateau border moving on a solid substrate at a velocity $V$.
The slip length $\lambda$ is small compared to the size of the Plateau border $r_{\rm pb}$, which confines the dissipation near the contact line.
This region can be described as a wedge of aspect ratio $h/L\ll 1$.
(e) Illustration of the case where the slip length is larger than the Plateau border.
In this case, the dissipation spans all over the Plateau border.
}
\label{fig:tau_vs_Ca}
\end{figure*}

The liquid deposition of a meniscus on a textured surface has been investigated experimentally and theoretically by Seiwert \textit{et al} \cite{seiwert_2011}.
For velocities such that the thickness predicted by Landau and Levich on a smooth surface is lower than the height of the pillars, the asperities are filled by the liquid, which means  there is no deposition on top of the pillars.
On foams, this condition writes $ {\rm Ca} < (b/r_{\rm pb})^{3/2}$, which is always satisfied in our experiments.
Consequently, we describe the dynamics of the Plateau borders with the Tanner description \cite{Tanner1979} and we take into account the presence of asperities filled with liquid by introducing an effective slip length $\lambda$.

We distinguish two limits according to the size of the slip length relative to the size of the Plateau border.
As schematized in Fig. \ref{fig:tau_vs_Ca}d, if $\lambda \ll r_{\rm pb} $, the
dissipation is mostly localized in the vicinity of the contact line where the shear rate is much larger than in the rest of the Plateau border.
This situation is essentially the one described by Tanner \cite{Tanner1979, Wei2019,Chan2020}, in which the liquid forms a wedge of height $h$ and length $L$, with a contact angle $\theta \sim h/L \ll 1$.
Thus, the capillary force per unit length $F_{\gamma} \sim \gamma \theta^2$ is balanced by the viscous force per unit length, which is proportional to the shear rate, $F_{\eta} \sim \eta \frac{V}{h^2} Lh \sim \eta \frac{V}{\theta}$.
This leads to the Tanner's law $\theta \sim {\rm Ca}^{1/3}$, and to a friction force $ F_{\eta}$ proportional to $\gamma {\rm Ca}^{2/3}$.
As a result, the predicted stress normalized by the value for a smooth surface (Eq.~\ref{eq:tau_smooth}) becomes independent of the capillary number
\begin{equation}
    \tau / \tau_{\rm s} \sim {\rm Ca}^0.
\end{equation}

Conversely, for $\lambda \gg r_{\rm pb}$, the liquid is sheared mainly over the slip length so that the dissipation occurs in the entire plateau border (Fig. \ref{fig:tau_vs_Ca}e).
This leads to a viscous force per unit length that scales as $\eta V r_{\rm pb}^2/\lambda^2 \sim \gamma (r_{\rm pb}/\lambda)^2 {\rm Ca}$.
Then, the resulting normalized stress writes
\begin{equation}
    \tau / \tau_{\rm s} \sim {\rm Ca}^{1/3}.
\end{equation}

We expect that the slip length is set mainly by the liquid height $b$ between the pillars.
The wetting condition to fulfill a self-SLIPS imposes that $b < r_{\rm pb}$.
Thus, the slip length must remain smaller than the Plateau border size in our experiments.
As shown in Fig. \ref{fig:tau_vs_Ca}b, the experimental values of the normalized stress for pillar heights $b$ between 10 and 80 $\mu$m are independent of the capillary number, which is in agreement with the prediction for  $\lambda \ll r_{\rm pb}$.
This conclusion is also corroborated by Keiser \textit{et al.}, who considered the sliding motion of a water drop on a surface infused with oil where the dissipation occurs in the oil foot of a shape comparable to a Plateau border \cite{Keiser2020}.

%
%

\section{Conclusions}

To summarize, drag reduction of liquids on solids has been recently achieved by infusing oil in textured surfaces.
However, former studies on the interaction of liquid foams with surfaces coated by an oil layer demonstrated either transport of oil by the foam or a foam destruction, preventing to use SLIPS to reduce friction on surfaces \cite{Mensire2017,Wong2021}.
The study of liquid foams interacting with textured surfaces  indicates that for asperities smaller than the size of the Plateau borders, the asperities are filled by liquid provided by the foam \cite{Commereuc2024}, offering a self-SLIPS.
We demonstrate experimentally that these surfaces effectively lower the friction compared to a smooth surface made of the same material.
Our measurements indicates that the depth of the asperities plays a significant role on the reduction of friction, which can reach up to about 30~\% independently of the foam velocity.
A major advantage of this design is the absence of any additional liquid on the surface, which enhances the maintenance of surface performance over time.

In this study, we focused on monodisperse foams.
This choice is motivated to limit the foam ageing and to maximize the reproducibility of the results, to offer an interesting avenue for further investigation.
To proceed towards the application, further studies will be necessary to examine the effects of foam polydispersity, high liquid fractions, high capillary numbers, and complex rheology. Additionally, the specific role of the texture geometry remains to be elucidated.
Indeed, measurements performed on surfaces decorated with randomly distributed glass beads has not revealed such drag reduction \cite{Marchand2020}, which is attributed to the limited film thickness caused by the bead spacing.
The present study can serve as a proof-of-concept to optimize the foam flow dissipation and develop a method suitable for a large scale production of textured surfaces.

\section*{Acknowledgments}
We kindly thank  Antoine Boury and Raphael Weil for their guidance on the soft photolithography.
We acknowledge for funding support from the French Agence Nationale de la Recherche in the framework of project AsperFoam - 19-CE30-0002-01.

\bibliography{biblio}

\providecommand{\noopsort}[1]{}\providecommand{\singleletter}[1]{#1}%
\begin{thebibliography}{10}

\bibitem{Barnes1995}
H.~A. Barnes.
\newblock A review of the slip (wall depletion) of polymer solutions, emulsions
  and particle suspensions in viscometers: its cause, character, and cure.
\newblock {\em Journal of Non-Newtonian Fluid Mechanics}, 56(3):221--251, 1995.

\bibitem{Kraynik1988}
A.~M. Kraynik.
\newblock Foam flows.
\newblock {\em Annu. Rev. Fluid Mech.}, 20(1):325--357, 1988.

\bibitem{Saugey2006}
A.~Saugey, W.~Drenckhan, and D.~Weaire.
\newblock Wall slip of bubbles in foams.
\newblock {\em Phys. Fluids}, 18(5):--, 2006.

\bibitem{Marze2008}
S.~Marze, D.~Langevin, and A.~Saint-Jalmes.
\newblock Aqueous foam slip and shear regimes determined by rheometry and
  multiple light scattering.
\newblock {\em J. Rheol.}, 52(5):1091--1111, 2008.

\bibitem{Nickerson2005}
C.~S. Nickerson and J.~A. Kornfield.
\newblock A “cleat” geometry for suppressing wall slip.
\newblock {\em Journal of Rheology}, 49(4):865--874, 2005.

\bibitem{Khan1988}
S.~A. Khan, C.~A. Schnepper, and R.~C. Armstrong.
\newblock Foam rheology: {III}. {M}easurement of shear flow properties.
\newblock {\em J Rheol}, 32(1):69--92, 1988.

\bibitem{Marchand2020}
M.~Marchand, F.~Restagno, E.~Rio, and F.~Boulogne.
\newblock Roughness-induced friction in liquid foams.
\newblock {\em Physical Review Letters}, 124:118003, 2020.

\bibitem{Wong2011}
T.-S. Wong, S.~H. Kang, S.~K.~Y. Tang, E.~J. Smythe, B.~D. Hatton, A.~Grinthal,
  and J.~Aizenberg.
\newblock Bioinspired self-repairing slippery surfaces with pressure-stable
  omniphobicity.
\newblock {\em Nature}, 477(7365):443--447, 2011.

\bibitem{Lafuma2011}
A.~Lafuma and D.~Qu{\'{e}}r{\'{e}}.
\newblock Slippery pre-suffused surfaces.
\newblock {\em Europhysics Letters}, 96(5):56001, nov 2011.

\bibitem{Hardt2022}
S.~Hardt and G.~McHale.
\newblock Flow and drop transport along liquid-infused surfaces.
\newblock {\em Annu. Rev. Fluid Mech.}, 54(1):83--104, 2022.

\bibitem{Heydarian2021}
S.~Heydarian, R.~Jafari, and G.~Momen.
\newblock Recent progress in the anti-icing performance of slippery
  liquid-infused surfaces.
\newblock {\em Progress in Organic Coatings}, 151:106096, 2021.

\bibitem{Epstein2012}
A.~K. Epstein, T.-S. Wong, R.~A. Belisle, E.~M. Boggs, and J.~Aizenberg.
\newblock Liquid-infused structured surfaces with exceptional anti-biofouling
  performance.
\newblock {\em Proceedings of the National Academy of Sciences},
  109(33):13182--13187, 2012.

\bibitem{Pakzad2022}
H.~Pakzad, A.~Nouri-Borujerdi, and A.~Moosavi.
\newblock Drag reduction ability of slippery liquid-infused surfaces: A review.
\newblock {\em Progress in Organic Coatings}, 170:106970, 2022.

\bibitem{Mensire2017}
R.~Mensire and E.~Lorenceau.
\newblock Stable oil-laden foams: Formation and evolution.
\newblock {\em Advances in Colloid and Interface Science}, 247:465--476, 2017.

\bibitem{Wong2021}
W.~S.~Y. Wong, A.~Naga, L.~Hauer, P.~Baumli, H.~Bauer, K.~I. Hegner,
  M.~D’Acunzi, A.~Kaltbeitzel, H.-J. Butt, and D.~Vollmer.
\newblock Super liquid repellent surfaces for anti-foaming and froth
  management.
\newblock {\em Nature Communications}, 12(1):5358, 2021.

\bibitem{Commereuc2024}
A.~Commereuc, S.~Mariot, E.~Rio, and F.~Boulogne.
\newblock Straight to zigzag transition of foam pseudo plateau borders on
  textured surfaces.
\newblock {\em Phys. Rev. Fluids}, 9:L041601, Apr 2024.

\bibitem{Drenckhan2015}
W.~Drenckhan and A.~Saint-Jalmes.
\newblock The science of foaming.
\newblock {\em Adv. Colloid Interface Sci.}, 222:228 -- 259, 2015.

\bibitem{Cantat2013b}
I.~Cantat, S.~Cohen-Addad, F.~Elias, F.~Graner, R.~H{\"o}hler, O.~Pitois,
  F.~Rouyer, A.~Saint-Jalmes, R.~Flatman, and S.~Cox.
\newblock {\em Foams: Structure and Dynamics}.
\newblock OUP Oxford, 2013.

\bibitem{Boulogne2023}
F.~Boulogne, E.~Rio, and F.~Restagno.
\newblock Evaporation-induced temperature gradient in a foam column.
\newblock {\em Langmuir}, 39(40):14256--14262, 2023.

\bibitem{Restagno2001}
F.~Restagno, J.~Crassous, E.~Charlaix, and M.~Monchanin.
\newblock A new capacitive sensor for displacement measurement in a
  surface-force apparatus.
\newblock {\em Meas. Sci. Technol.}, 12(1):16, 2001.

\bibitem{Bretherton1961}
F.~P. Bretherton.
\newblock The motion of long bubbles in tubes.
\newblock {\em J. Fluid Mech.}, 10:166--188, 3 1961.

\bibitem{Hirasaki1985}
G.J. Hirasaki and J.B. Lawson.
\newblock Mechanisms of foam flow in porous media: apparent viscosity in smooth
  capillaries.
\newblock {\em Soc Petrol Eng J}, 25(02):176--190, 1985.

\bibitem{Cantat2013a}
I.~Cantat.
\newblock Liquid meniscus friction on a wet plate: Bubbles, lamellae, and
  foams.
\newblock {\em Phys. Fluids}, 25(3):031303, 2013.

\bibitem{Raufaste2009}
C.~Raufaste, A.~Foulon, and B.~Dollet.
\newblock Dissipation in quasi-two-dimensional flowing foams.
\newblock {\em Phys. Fluids}, 21(5):053102, 2009.

\bibitem{Emile2013}
J.~Emile, H.~Tabuteau, F.~Casanova, and O.~Emile.
\newblock Bubble-wall friction in a circular tube.
\newblock {\em Soft Matter}, 9:4142--4144, 2013.

\bibitem{seiwert_2011}
J.~Seiwert, C.~Clanet, and D.~Quéré.
\newblock Coating of a textured solid.
\newblock {\em Journal of Fluid Mechanics}, 669:55--63, 2011.

\bibitem{Tanner1979}
L.~H. Tanner.
\newblock The spreading of silicone oil drops on horizontal surfaces.
\newblock {\em Journal of Physics D: Applied Physics}, 12(9):1473, sep 1979.

\bibitem{Wei2019}
H.-H. Wei, H.-K. Tsao, and K.-C. Chu.
\newblock Slipping moving contact lines: critical roles of de {G}ennes’s
  ‘foot’ in dynamic wetting.
\newblock {\em Journal of Fluid Mechanics}, 873:110–150, 2019.

\bibitem{Chan2020}
T.~S. Chan, C.~Kamal, J.~H. Snoeijer, J.~E. Sprittles, and J.~Eggers.
\newblock Cox–voinov theory with slip.
\newblock {\em J. Fluid Mech.}, 900:A8, 2020.

\bibitem{Keiser2020}
A.~Keiser, P.~Baumli, D.~Vollmer, and D.~Qu\'er\'e.
\newblock Universality of friction laws on liquid-infused materials.
\newblock {\em Phys. Rev. Fluids}, 5:014005, Jan 2020.

\end{thebibliography}

\bibliographystyle{unsrt}


\end{document}